\documentclass[preprint]{revtex4}
     
\usepackage{graphicx}
\usepackage{dcolumn}
\usepackage{bm}
\usepackage{txfonts}
\usepackage{pifont}
\usepackage{amssymb}

\begin{document}

\bibliographystyle{naturemag}

\title{Quantum criticality in a metallic spin liquid}

\author{Y. Tokiwa$^1$}
\author{J. J. Ishikawa$^2$}
\author{S. Nakatsuji$^{2,3}$}
\author{P. Gegenwart$^1$}

\affiliation{$^1$I. Physikalisches Institut, Georg-August-Universit\"{a}t G\"{o}ttingen, 37077 G\"{o}ttingen, Germany}
\affiliation{$^2$Institute for Solid State Physics, University of Tokyo, Kashiwa 277-8581, Japan}
\affiliation{$^3$PRESTO, Japan Science and Technology Agency (JST), 4-1-8 Honcho Kawaguchi,
Saitama 332-0012, Japan}

\begin{abstract}
When magnetic order is suppressed by frustrated interactions, spins form a highly correlated fluctuating "spin liquid" state  down to low temperatures. Magnetic order of local moments can also be suppressed when they are fully screened by conduction electrons through the Kondo effect. Thus, the combination of strong geometrical frustration and Kondo screening may lead to novel types of quantum phase transitions~\cite{Si-PhysicaB06,Vojta-prb08,custers-prl10}. We report low-temperature thermodynamic measurements on the frustrated Kondo lattice Pr$_2$Ir$_2$O$_7$~\cite{Nakatsuji-prl06,Machida-Nature10}, which displays a chiral spin liquid state below 1.5 K due to the frustrated interaction between Ising 4f local moments and their interplay with Ir conduction electrons. Our results provide a first clear example of zero-field quantum critical scaling that emerges in a spin liquid state of a highly frustrated metal.
\end{abstract}

\maketitle

The relation between spin liquid and quantum critical states is illustrated in Figure 1. Here the parameter $Q$ is a measure of the strength of quantum fluctuations, induced by competing interactions or geometrical frustration. For insulators it could be quantified by the frustration parameter $f=\left|\theta\right|/T_{\rm MO}$ which measures the ratio of the absolute value of the Weiss temperature to the ordering temperature. Insulators with sufficiently large frustration of their local magnetic moments form a strongly correlated fluctuating state, so-called spin liquid, at temperatures below $\left|\theta\right|$. For the fully frustrated case the ordering temperature is completely suppressed, giving rise to a quantum critical point (QCP) and the spin liquid phase extends down to zero temperature. On the other hand, Kondo screening in metals acts in reducing the size of local magnetic moments and suppresses magnetic order as well. As a result, in the presence of Kondo interaction, already a moderately strong frustration can induce quantum criticality by suppression of magnetic order to zero temperature.  A line of quantum critical points emerges (cf. red line in Fig. 1) which connects at $T=0$ the insulating spin liquid with the metallic sector of the phase diagram. Different types of paramagnetic states are currently discussed for f-electron based heavy-fermion metals. For sufficiently large Kondo coupling $K$ the f-moments are bound in Kondo singlets. The resulting Fermi liquid (FL) has a "large" Fermi surface volume including the f electrons. However, the possibility of an f-selective Mott transition due to a breakdown of Kondo screening has extensively been discussed in recent years~\cite{ColemanP:HowdFl,SiQM:Loccqp,paschen}. It is expected, that for sufficiently strong $Q$ a novel paramagnetic phase $P_S$ forms, with "small" Fermi volume determined by conduction electrons only~\cite{Senthil-prl03,Si:2013,custers-prl10}. Since the moments in this state are neither Kondo screened nor magnetically ordered, they are expected to form a spin-liquid. Experimentally this phase is largely unexplored and its relation to the proposed line of QCPs connecting to the insulating spin liquid is unknown.

We focus on the pyrochlore iridate Pr$_2$Ir$_2$O$_7$, with local Pr$^{3+}$ magnetic moments and a small concentration of Ir 5d conduction electrons giving rise to a weak Kondo coupling~\cite{Nakatsuji-prl06}. Crystal electric field (CEF) splitting leads to a magnetic non-Kramers doublet ground state and the lowest CEF excited state at 162\,K~\cite{Nakatsuji-prl06}. Those moments are located on the edges of corner-sharing tetrahedra, leading to a geometrically frustrated magnetic interaction. The effect of strong geometrical frustration is evidenced by the absence of magnetic ordering upon cooling the system to very low temperatures, despite a Curie Weiss temperature $\Theta=-20$ K, determined from the temperature dependence of the susceptibility above 100\,K~\cite{Nakatsuji-prl06} and the estimated nearest neighbor ferromagnetic interaction of 1.4\,K~\cite{Machida-Nature10}. The metamagnetic transition only observed along the [111] axis confirms the spin ice formation while the transition field of 2.3\,T yields a ferromagnetic interaction scale of 1.4\,K. The combination of a finite Kondo coupling $K$ and the strong geometrical frustration places Pr$_2$Ir$_2$O$_7$ in the $P_S$ phase of Fig.~1. An intriguing observation in this system is the spontaneous Hall conductivity below 1.5\,K in the absence of magnetic order~\cite{Machida-Nature10}. This was attributed to a formation of a chiral spin liquid. Here the non-zero spin chirality acts as a fictitious field and breaks time reversal symmetry. This is distinct from the topological Hall signal due to skyrmions and its related non-Fermi liquid phase in MnSi~\cite{ritz-nature13}.

The manifold of degenerate configurations of spin ice systems results in the Pauling entropy ($R$/2)ln(3/2)~\cite{Ramirez-Nature99}. Classical spin ice systems, such as Dy$_2$Ti$_2$O$_7$~\cite{Ramirez-Nature99} and Ho$_2$Ti$_2$O$_7$~\cite{Bramwell-prl01} develop two-in two-out configurations from a completely uncorrelated state at high temperatures. Their specific heat, $C(T)$, exhibits a characteristic maximum upon cooling, releasing the entropy of $R[\ln(2)-(1/2)\ln(3/2)]$. Qualitatively similar behavior has also been found for insulating Pr$_2$Zr$_2$O$_7$~\cite{Kimura-NatComm13}. The ground state of spin ice is controversially discussed~\cite{pomaranskiabsence2013}. Pr$_2$Ir$_2$O$_7$ satisfies the requirements of spin ice formation, namely Ising type 4$f$ magnetic moments and ferromagnetic nearest neighbor interaction. However, the presence of conduction electrons gives rise to a RKKY-type interaction which has been proposed to mediate spin chirality~\cite{Flint-prb13} or spiral dipolar and XY-quadrupolar ordering~\cite{Lee-arxiv13}. We have studied the thermodynamic properties of Pr$_2$Ir$_2$O$_7$ down to below 100\,mK.

Figure 2 displays our specific heat and entropy data on single crystalline Pr$_2$Ir$_2$O$_7$. Upon cooling the specific heat passes the characteristic spin ice maximum and decreases until it reaches a minimum at 0.4\,K. At this temperature the entropy approaches the Pauling value, indicating that the formation of two-in two-out spin-ice configurations is completed in all of the tetrahedra. Further cooling below 0.4\,K yields to a huge enhancement of the specific heat divided by temperature (see inset Fig.~2). In the absence of a clear phase transition anomaly we associate the corresponding entropy reduction to the melting of spin ice configurations by quantum fluctuations at $T_m\approx 0.4$~K (see arrow in Fig.~2)~\cite{Onoda-prl10}.

Next, we turn our attention to the detection of quantum criticality in this material. The magnetic  Gr\"{u}neisen ratio $\Gamma_H=1/T (dT/dH)_S$, which measures the change of temperature with magnetic field under adiabatic conditions generically diverges in the approach of any QCP~\cite{zhu}. This results from the entropy accumulation in the quantum critical regime and has experimentally been confirmed e.g. at the field-tuned QCP in the heavy-fermion metal YbRh$_2$Si$_2$~\cite{tokiwa-prl09}. Assuming that critical behavior is governed by a single diverging time scale near the QCP, universal scaling of physical properties like thermal expansion, specific heat or magnetization as a function of $T/(H-H_c)^{\nu z}$, where $\nu$ is the correlation length exponent and $z$ is the dynamical exponent, is expected~\cite{zhu}. Such scaling results from the competition of quantum and thermal fluctuations near the QCP. The distance to QCP, $H-H_c$, which is related to quantum fluctuations, influences the critical scaling, in contrast to the classical critical scaling, where the critical components of physical quantities depend only on the reduced temperature, $\left|T-T_c\right|$/$T_c$ ($T_c$; critical temperature). Observation of $T/(H-H_c)^{\nu z}$ scaling over a substantial parameter range therefore acts as proof of quantum critical behavior and determines the exact position of the QCP, i.e. the value of the critical magnetic field $H_c$. In Sr$_3$Ru$_2$O$_7$~\cite{gegenwart-prl06}, CeCu$_6$~\cite{schroeder} and YbRh$_2$Si$_2$~\cite{custers}, several physical quantities were shown to exhibit quantum critical scaling in a wide range in $H$-$T$ phase space around their QCPs.

The magnetic Gr\"uneisen ratio can also be expressed by the magnetization $M$ and specific heat $C$, as $\Gamma_H=-(dM/dT)/C$~\cite{zhu}. Since for Pr$_2$Ir$_2$O$_7$ the low-temperature magnetization is almost isotropic at fields below about 0.5~T~\cite{Machida-Nature10}, we do not expect that the zero-field quantum critical behavior of the magnetic Gr\"uneisen ratio discussed in the following displays an anisotropy with respect to the field orientation at low fields. Figure~3 displays $\Gamma_H/H$ of Pr$_2$Ir$_2$O$_7$ at various magnetic fields applied parallel to the [111] direction, while data for the field along [100] are shown in supplemental material (SM). Upon cooling, the magnetic Gr\"uneisen ratio at low field diverges according to $T^{-3/2}$ over almost one decade in temperature down to about 0.4~K, providing evidence for quantum critical behavior~\cite{zhu}. Remarkably, this temperature coincides with the reduction of the entropy below the Pauling value ($T_m$ in Fig.2). Whatever causes the "melting" of the spin-ice state, it also suppresses quantum criticality. As the magnetic field is increased, $\Gamma_H/H$ becomes temperature independent at lowest temperatures, indicating the formation of a quantum paramagnetic state, i.e. a non-critical quantum spin liquid. Assuming universal scaling near a QCP~\cite{zhu}, it is theoretically expected that $\Gamma_H=-G_r/(H-H_c)$ in the quantum paramagnetic state. The universal prefactor $G_r=\nu(d-y_0z)/y_0$ is determined by the dimensionality $d$ and the exponent $y_0$ of the temperature dependence of the specific heat ($C\sim T^{y_0}$). For Pr$_2$Ir$_2$O$_7$ the large uncertainty in the subtraction of the nuclear specific heat makes it impossible to reliably determine $y_0$. The constant $\Gamma_H$ values obtained at lowest temperatures follow the expected $(H-H_c)^{-1}$ dependence and are remarkably well fitted by $H_c$=0$\pm$0.04\,T, suggesting a zero-field QCP (Fig.~3 inset). As explained in SM, slightly different scaling relations are expected in such case, because of the quadratic coupling of $H$ with order parameter. As a result, the Gr\"uneisen ratio $\Gamma_HH=f(T/H^{2\nu z})$, where $f$ is a universal scaling function. Furthermore, for $T\rightarrow 0$, the magnetic Gr\"{u}neisen ratio diverges toward zero field as $\Gamma_H=-2G_r/H$ in case of $H_c$=0. Experimentally, we found $2G_r=-0.25\pm 0.03$ for $H\parallel$[111].
A slight deviation from the $\Gamma_H=-2G_r/H$ scaling, is found above 2~T, upon approaching the transition to the 3-in 1-out state at 2.3\,T~\cite{Machida-Nature10} which breaks the ice-rule. Thus, quantum criticality is related to the degeneracy within the spin-ice state.

We further examine quantum criticality in Pr$_2$Ir$_2$O$_7$ by rescaling the $\Gamma_H$ data. In Figure 4 the magnetic Gr\"uneisen ratio data are displayed as $\Gamma_HH$ vs $T$/$H^{4/3}$. The measured data collapse on a common curve for about four decades in the $x$- and more than three decades in the $y$-axis. This confirms that the system is located at a zero-field QCP. The scaling plot clearly shows the crossover between the quantum critical and quantum paramagnetic states, that are characterized by $\Gamma_H\sim HT^{-3/2}$ and $\Gamma_H=0.25/H$ (temperature independent), respectively (cf. also inset of Fig.3). There is a certain regime in temperature-field phase space for which scaling does not hold. At low fields, this includes temperatures below 0.4~K and this extends up to about 0.35~T, while at larger fields all data follow scaling down to lowest temperatures. The same scaling behavior of $\Gamma_HH$ vs $HT^{-3/2}$ is found also for $H\parallel$[100] (see SM). However, the value of 2$G_r$=-0.37$\pm$0.03 in the quantum paramagnetic state at low temperatures is different from -0.25 for $H\parallel$[111]. As the field direction of [111] is known as a special direction, stabilizing the kagome-ice state~\cite{Hiroi-jpsj03}, the anisotropy of $G_r$ may be a result of quantum paramagnetic states with different $y_0$ along the two field directions. Generically, as found e.g. for the QCP in CeCoIn$_5$~\cite{tokiwa-prl13}, $G_r$ is isotropic (see discussion in SM).

Figure 5 displays the $T-H$ phase diagram, where the color coding indicates the size of the magnetic Gr\"uneisen ratio which displays quantum critical scaling for temperatures below 3~K and fields below 2~T. Only very close to the zero-field QCP this scaling is cut-off in the low-temperature and low-field regime bound by the blue dotted line. In the absence of a clear phase transition the nature of this state is unknown.

The experimentally observed scaling exponents allow to characterize quantum criticality in Pr$_2$Ir$_2$O$_7$. If quantum criticality is governed by a single diverging time scale~\cite{zhu}, the magnetic Gr\"uneisen ratio scales like $T/(H)^{2\nu z}$, yielding $\nu z$=2/3. This differs from the expectation within the itinerant Hertz-Millis-Moriya theory for magnetic QCPs. The latter predicts $\nu z$=1 for antiferromagnetism and $\nu z$=3/2 for ferromagnetism~\cite{zhu}. Since our system has a local moment character and only very weak Kondo interaction, it is not surprising that the observed exponents differ from the itinerant model for spin-density-wave instabilities. Furthermore, within the quantum paramagnetic state we observe $\Gamma_HH=$0.25 and 0.37, implying $-\nu(d-y_0z)/y_0$=0.125 and 0.185 for $H\parallel$[111] and [100], respectively. 
These relations will place strong constraints on theoretical modeling of quantum criticality in Pr$_2$Ir$_2$O$_7$~\cite{Moon-arxiv13}.

The previous observation of a spontaneous Hall effect, despite the absence of dipolar magnetic order, suggests a chiral spin liquid state in the  temperature range~\cite{Machida-Nature10}, where our study reveals zero-field quantum criticality. Taken together, these observations support a novel type of "quantum critical spin liquid" state that may be expected from the global phase diagram of Kondo lattice materials. The observed quantum critical scaling is different from what has been predicted for itinerant systems with large Fermi volume suggesting our system to be placed within the "$P_S$" sector in the phase diagram. Several novel quantum phases like unconventional superconductivity~\cite{mathur:nature-98} or electronic nematic ordering~\cite{Borzi-science07} have been found in clean metals near quantum critical points. In this respect the state below 0.4\,K in Pr$_2$Ir$_2$O$_7$ requires further attention.

\newpage

{\bf Methods}

Single crystals of Pr$_2$I$_2$O$_7$ were grown using a flux method. The magnetocaloric effect and specific heat were measured respectively by utilizing the alternating field technique~\cite{tokiwa-rsi11} and the quasi-adiabatic heat pulse method in a dilution refrigerator, equipped with a superconducting magnet. Specific heat measurements in the high temperature region above 2\,K were performed utilizing a commercial Quantum Design Physical Property Measurement System. 

{\bf Acknowledgments}

The authors acknowledge discussions with L. Balents, M. Brando, J.G. Donath, M. Garst, Yong-Baek Kim, K. Kimura, Q. Si, C. Stingl, M. Vojta and K. Winzer. This work has been supported by the German Science Foundation through FOR 960 (Quantum phase transitions), the Helmholtz Virtual Institute VH521, and by Grants-in-Aid for Scientific Research (No. 25707030) from JSPS, and by PRESTO of JST. The use of the Materials Design and Characterization Laboratory at ISSP is gratefully acknowledged. This work was supported also in part by the National Science Foundation under Grant No. PHYS-1066293 and the hospitality of the Aspen Center for Physics.

{\bf Correspondence}

Correspondence and requests for materials should be addressed to Y.T.~(email: ytokiwa@gwdg.de) or P.G.~(email: pgegenw@gwdg.de).

\bibliographystyle{nature}

\begin{thebibliography}{10}
\bibitem{note-arxiv13}
\bibinfo{note}{Very recently a novel non-Fermi liquid phase has been proposed
  for itinerant pyrochlore iridates$^{29}$. However this model
  does not take into account magnetic entropy from Pr moments present in our
  system.}
\end{thebibliography}

\newpage

\begin{figure}[t]
\includegraphics[width=\linewidth,keepaspectratio]{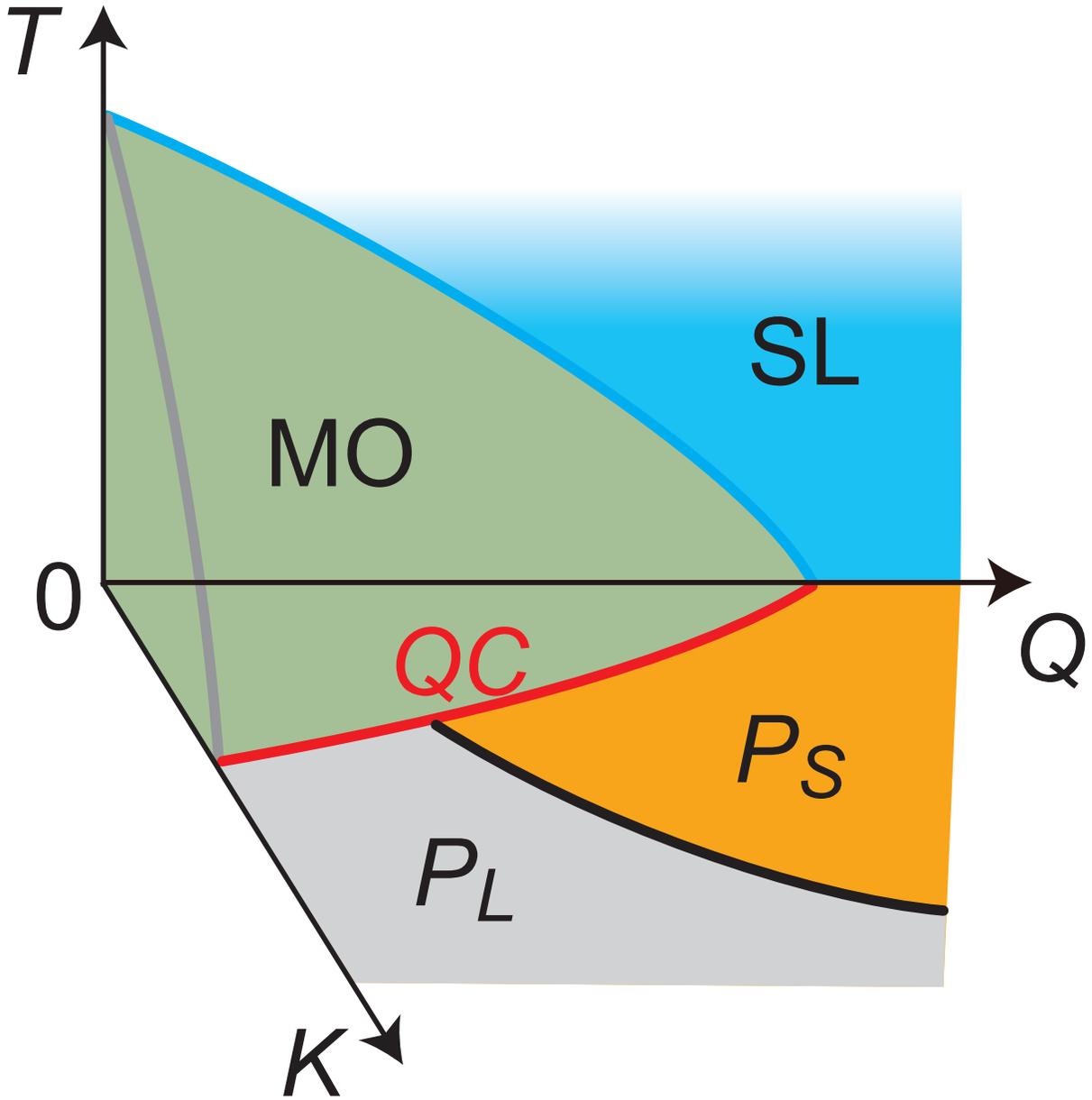}
\caption{{\bf Schematic phase diagram for geometrically frustrated Kondo-lattice systems.} Different phases are shown as function of temperature ($T$), strength of quantum fluctuations ($Q$) and Kondo-coupling ($K$). The blue phase in the sector without Kondo coupling indicates an insulating spin liquid (SL). A line of quantum critical points (in red) bounds the long-range magnetic order (MO) (green region) at $T=0$. The black line in the $T=0$ plane indicates the transition between paramagnetic states with large (P$_L$) and small (P$_S$) Fermi surface, whose continuation within the MO state has been omitted for clarity~\cite{Si:2013,custers-prl10}. The proposed and yet unexplored P$_S$ phase comprises unscreened local moments in a metallic spin liquid state~\cite{Senthil-prl03,Vojta-prb08}.}
\end{figure}

\begin{figure}[t]
\includegraphics[width=0.6\linewidth,keepaspectratio]{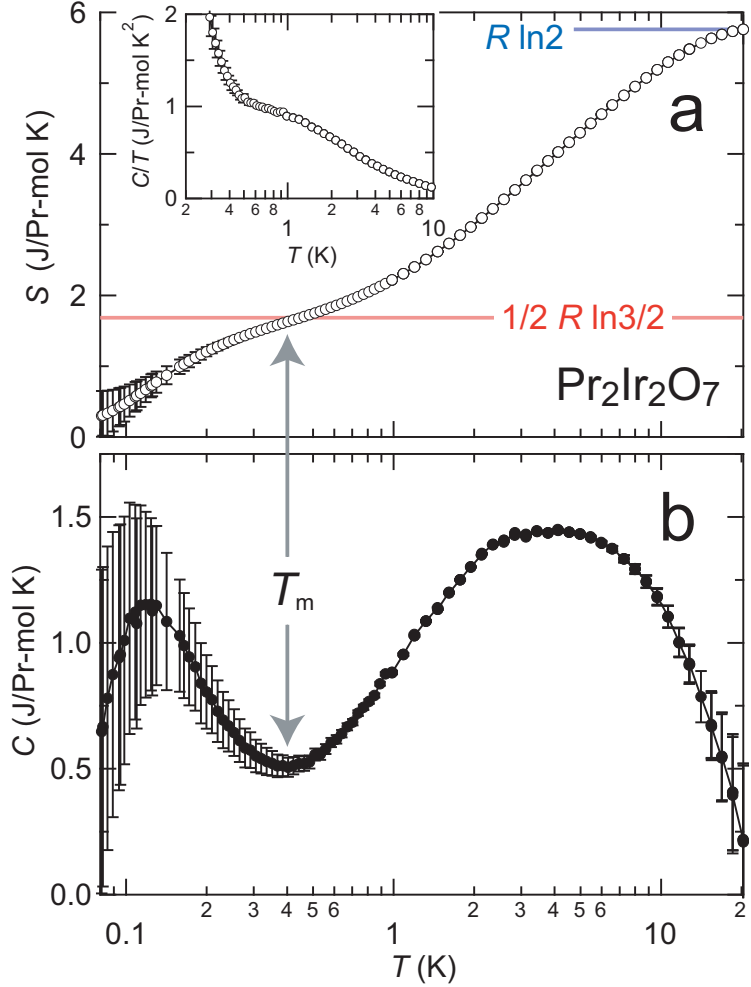}
\caption{(Color online) {\bf Low-temperature specific heat and entropy of Pr$_2$I$_2$O$_7$.} (a) Magnetic entropy calculated from the specific heat data of (b) under the assumption that the entropy at 20~K approaches the value expected for magnetic doublets R$ln(2)$. This is justified, since the first excited crystalline electric field state is well separated by 162\,K from the ground state~\cite{Nakatsuji-prl06}. Entropy for the crystalline electric field doublet, $R$ln(2), and Pauling entropy, ($R$/2)ln(3/2), are marked by blue and and red lines, respectively. (b) Specific heat $C$ as function of temperature. Phonon contribution (determined on non-magnetic reference Er$_2$Ir$_2$O$_7$), contribution due to the excited crystalline electric field states, as well as, nuclear contribution have been subtracted (see  supplemental material). The large error bars at low temperatures are caused by the uncertainty of the subtraction of the nuclear contribution. Onset of low-temperature upturn in $C(T)$, indicative of the quantum melting of spin ice at $T_m$=0.4\,K is marked by the arrows. (a)  Inset displays the respective specific heat divided by temperature, $C/T$, on a logarithmic temperature scale.}
\end{figure}

\begin{figure}[t]
\includegraphics[width=\linewidth,keepaspectratio]{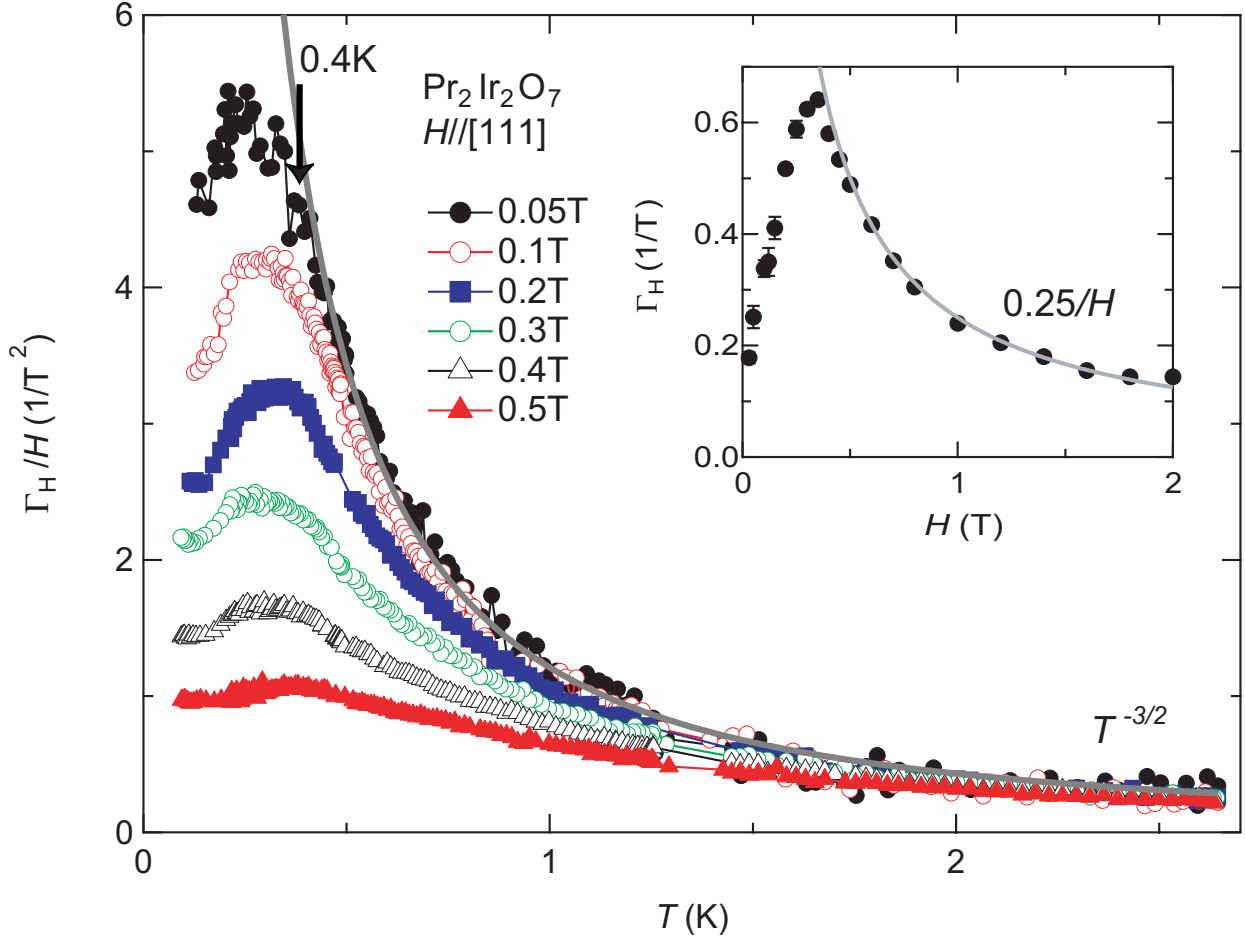}
\caption{{\bf Divergent behavior of the magnetic Gr\"uneisen ratio, $\Gamma_{\rm H}$, of Pr$_2$Ir$_2$O$_7$ as function of temperature.} $\Gamma_{\rm H}$ is measured at magnetic fields applied parallel to the [111] direction.  The divergence with a power law $\sim T^{-3/2}$ is indicated by the gray solid line. Arrow at 0.4~K indicates the temperature, below which $\Gamma_{\rm H}$ deviates from the power-law divergence. Inset shows the field dependence of the constant $\Gamma_H$ values in the low-temperature quantum paramagnetic regime at $T\rightarrow 0$. Gray solid line indicates 0.25/$H$.}
\end{figure}

\begin{figure}[t]
\includegraphics[width=\linewidth,keepaspectratio]{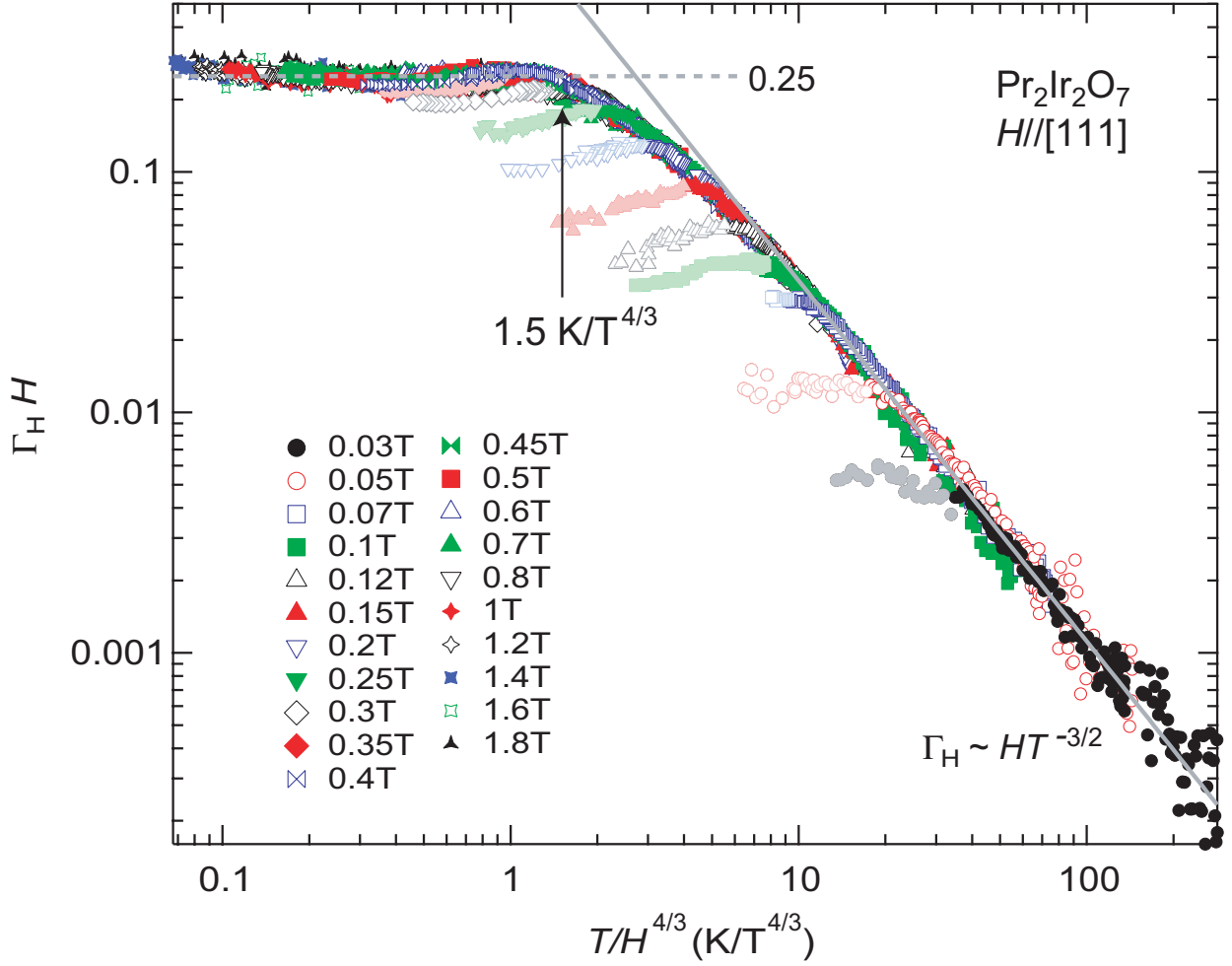}
\caption{{\bf Evidence of a zero-field quantum critical point in Pr$_2$Ir$_2$O$_7$.} Quantum critical scaling of magnetic Gr\"uneisen ratio, $\Gamma_{\rm H}$, is shown by plotting $\Gamma_{\rm H}H$ vs $T/H^{4/3}$. Magnetic field is applied parallel to the [111] direction. Data points from the phase space regime indicated by gray dotted line in Fig. 5, that deviate from the scaling behavior, are plotted in pastel colors. Constant $\Gamma_HH$=0.25 in the quantum paramagnetic state is indicated by gray dotted line. The arrow at 1.5\,$T/H^{4/3}$ indicates the crossover between the quantum paramagnetic and quantum critical behavior. Solid gray line indicates $HT^{-3/2}$ dependence of $\Gamma_{\rm H}$ within the quantum critical state.}
\end{figure}

\begin{figure}[t]
\includegraphics[width=\linewidth,keepaspectratio]{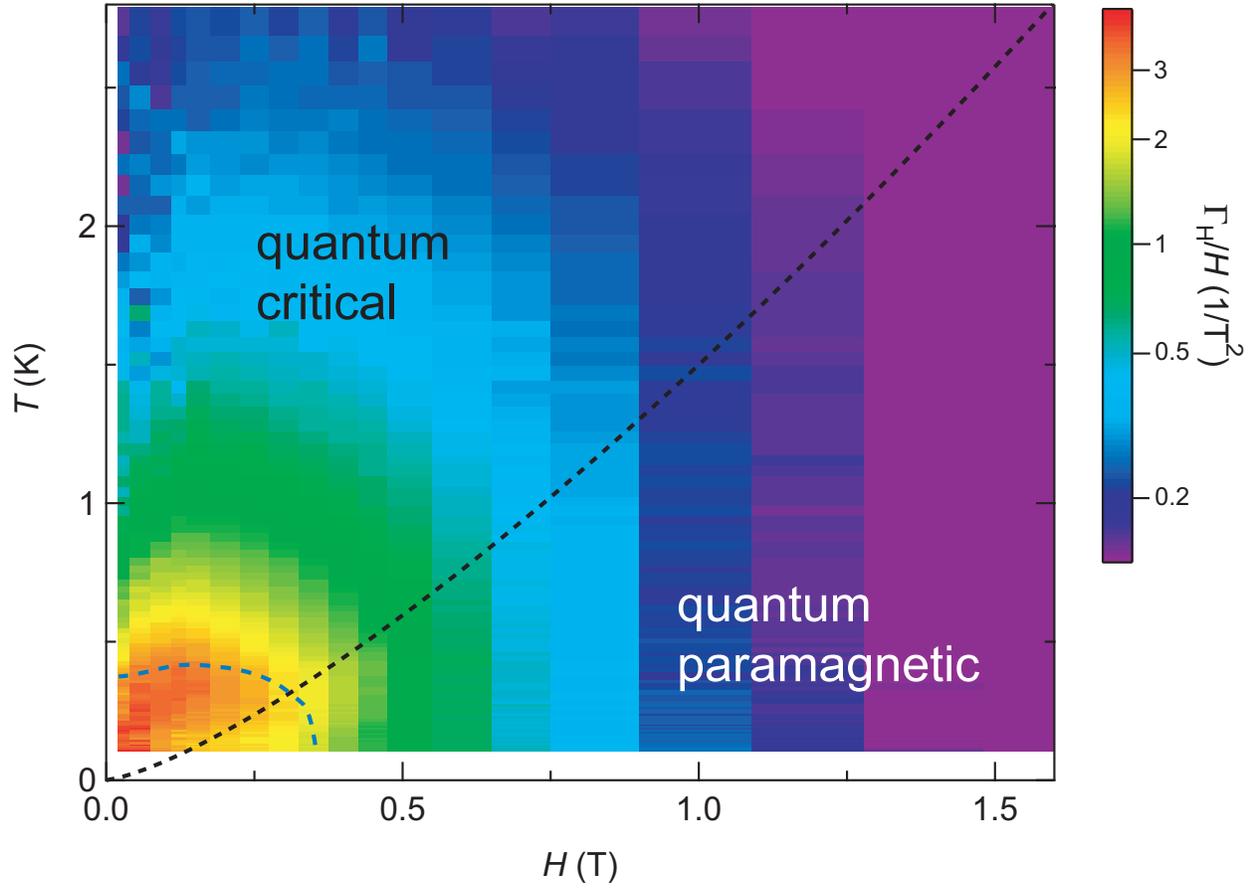}
\caption{{\bf Color-coded contour plot of the Gr\"uneisen ratio divided by magnetic field, $\Gamma_{\rm H}/H$, of Pr$_2$Ir$_2$O$_7$ in $H$-$T$ phase space.} Note the logarithmic scale for the color-coding. The crossover between the quantum critical and quantum paramagnetic regimes follows a $H^{4/3}$ dependence and is indicated by the black dotted line. The gray dotted lines encloses the phase space for which the magnetic Gr\"uneisen ratio deviates from the scaling behavior (cf. data points in pastel color in Fig. 4).}
\end{figure}

\end{document}